\documentclass[12pt]{iopart}
\usepackage{iopams}
\usepackage{amsmath}
\usepackage{amssymb}
\usepackage{amsthm}
\usepackage{cite}
\usepackage[urlcolor=blue,colorlinks=true,linkcolor=blue,pdfstartview={FitH},pdfcenterwindow=true,bookmarks=false]{hyperref}
\newcommand{\ket}[1]{\ensuremath{|#1\rangle}}
\newcommand{\bra}[1]{\ensuremath{\langle#1|}}

\newcommand{\eg}{\emph{e.g.}}
\newcommand{\ie}{\emph{i.e.}}
\newcommand{\etal}{\emph{et al}}
\newcommand{\mf}{\mathbf}

\newcommand{\mc}{\mathcal}
\newcommand{\mb}{\mathbb}
\newcommand{\ot}{\otimes}

\newcommand{\Id}[1]{\mb{I}_{\text{#1}}}

\newcommand{\p}{{\scriptscriptstyle{+}}}
\newcommand{\m}{{\scriptscriptstyle{-}}}
\newcommand{\spm}{{\scriptscriptstyle{\pm}}}

\newcommand{\dg}{\dagger}
\newcommand{\ani}[1]{a(#1)}
\newcommand{\cre}[1]{a^{\dg}(#1)}
\newcommand{\txt}[1]{\text{#1}}
\newcommand{\parity}{\txt{P}_{\pi}}

\begin{document}

\title[\it Spin-Boson model]{Exact solution of the  Schr\"{o}dinger equation with the spin-boson Hamiltonian}
%\title[\it Spin-Boson model]{Spin-boson model: an exact solution of the Schr\"{o}dinger equation}

\author{Bart{\l}omiej Gardas}

%\address{Institute of Theoretical and Applied Informatics, Polish Academy
%of Sciences, Ba{\l}tycka 5, 44-100 Gliwice, Poland}
%\vspace{2mm}
\address{Institute of Physics, University of Silesia, PL-40-007 Katowice, Poland}

\ead{bartek.gardas@gmail.com}

\begin{abstract}
We address the problem of obtaining the exact reduced dynamics of the spin-half (qubit)
immersed within the bosonic bath (enviroment). An exact solution of the Schr\"{o}dinger
equation with the paradigmatic spin-boson Hamiltonian is obtained. We believe that this
result is a major step ahead and may ultimately contribute to the complete resolution of
the problem in question. We also construct the constant of motion for the spin-boson system.
In contrast to the standard techniques available within the framework of the open quantum
systems theory, our analysis is based on the theory of block operator matrices.
\end{abstract}

%Uncomment for PACS numbers title message
% Keywords required only for MST, PB, PMB, PM, JOA, JOB?
%\vspace{2pc}
%\noindent{\it Keywords}: Article preparation, IOP journalsb
% Uncomment for Submitted to journal title message
%\submitto{\JPA}
% Comment out if separate title page not required
\maketitle

\section{Introduction}

The Hamiltonian of the paradigmatic spin-boson (SB) model is specified as~\cite{SB_Fannes, SB_Fannes2,SB_Spohn,SB_Honegger}

\begin{equation}
\label{eq:SB}
\mf{H}_{\text{SB}} = \txt{H}_{\text{S}}\ot\Id{B} + \Id{S}\ot \txt{H}_{\text{B}} + \mf{H}_{\text{int}},
\end{equation}
where

\begin{equation}
\label{eq:S}
\text{H}_{\text{S}} = (\beta\sigma_z+\alpha\sigma_x)
\quad\text{and}\quad
\txt{H}_{\text{B}} = \int_{0}^{\infty}d\omega\, h(\omega)\,\cre{\omega}\ani{\omega},
\end{equation}
are the Hamiltonian of the spin-half (qubit) and the bosonic field (environment), respectively.  The interaction between the systems
has the following form

\begin{equation}
\label{eq:CSB}
\mf{H}_{\text{int}} = \sigma_z\ot\int_{0}^{\infty}d\omega \left(g(\omega)^{*}\ani{\omega}+g(\omega)\cre{\omega}\right)
 \equiv\sigma_z\ot\txt{V}.
\end{equation}
$\Id{S}$ and $\Id{B}$ are the identity operators in corresponding Hilbert spaces of the qubit and the environment, respectively.

In the above description, $\sigma_z$ and $\sigma_x$ are the standard Pauli matrices. The bosonic creation $\cre{\omega}$ and annihilation
$\ani{\omega}$ operators obey the canonical commutation relation: $[\ani{\omega},\cre{\eta}]=\delta(\omega-\eta)\Id{B}$, for $\omega,
\eta\in[0,\infty)$. The functions $h$, $g\in L^{2}[0,\infty]$ model the energy of the free % (\eg, for photons or phonons $h(\omega)=\omega$)
bosons and the coupling the bosons with the qubit, respectively. The constants $\alpha$ and $\beta$ are assumed to be real and non
negative numbers. Furthermore, $\beta$ represents the energy gap between the eigenstates $\ket{0}$ and $\ket{1}$ of $\sigma_z$, while
$\alpha$ is responsible for the tunneling between these states. The Hamiltonian~\eref{eq:SB} acts on the total Hilbert space
$\mc{H}_{\text{tot}}=\mb{C}^{2}\ot\mc{F}_{\text{B}}$, where $\mc{F}_{\text{B}}:=\mc{F}(L^{2}[0,\infty])$ is the bosonic Fock space~\cite{SB_Alicki}.

It is worth mentioning that more often we encounter situations in which there is a countable number (finite, in particular) of bosons
(see \eg{}~\cite{Leggett,SB_work,SB_paradigm,SB_parity}). In such cases we define the SB model via the following Hamiltonian

\begin{equation}
\label{eq:DSB}
\mf{H}_{\text{SB}} = (\beta\sigma_z+\alpha\sigma_x)\ot\Id{B} +
               \Id{S}\ot\sum_{k}h_ka_k^{\dg}a_k
                   +
               \sigma_z\ot\sum_{k}\left(g_k^{*}a_k+g_ka^{\dg}\right),
\end{equation}
where the creation and annihilation operators $a_k^{\dg}$, $a_k$ satisfy $[a_k,a_{l}^{\dg}]=\delta_{kl}$.
Formally, it is possible to obtain~\eref{eq:DSB} from~\eref{eq:SB} by setting

\begin{equation}
 x(\omega) = \sum_{k}x_k\delta(\omega-\omega_k),\quad\txt{where}\quad x=h,g.
\end{equation}
Therefore, we can treat both cases simultaneously. Although generalizations of the SB model (\eg{} asymmetric coupling~\cite{dajka})
are also under intensive investigation, we will not focus on them in this paper.

The problem of a small quantum system coupled to the external degrees of freedom plays an important role in various
fields of modern quantum physics. The SB model provides a simple mathematical description of such coupling in the case
of two-level quantum systems. For instance, an interaction between two-level atoms and the electromagnetic radiation
can be modeled via the SB Hamiltonian~\cite{puri}. For this reason the SB model is of great importance to the modern
quantum optics. There are various physical problems (\eg, decoherence~\cite{zurek,dajka_cat,dajka1,dajka2}, geometric
phase~\cite{dajka_phase}) related to the properties of the model in question, which has already been addressed and intensively
discussed. Nonetheless, an exact solution of the Schr\"{o}dinger equation:

\begin{equation}
\label{eq:Sch}
 \rmi\partial_t\ket{\Psi_t} = \mf{H}_{\text{SB}}\ket{\Psi_t}
 \quad\text{with}\quad \ket{\Psi_0}\equiv\ket{\Psi},
\end{equation}
is still missing for both $\alpha\not=0$ and $\beta\not=0$. Several approximation methods~\cite{davies} have been developed
in past fifty years to manage this problem. Models obtained from the SB Hamiltonian under mentioned approximations are
well-established and in most cases they are exactly solvable. The famous Jaynes-Cummings model~\cite{GJC} can serve as an example.
Formally, one can always express the solution of~(\ref{eq:Sch}) as $\ket{\Psi_t}=\mf{U}_t\ket{\Psi}$, where 
$\mf{U}_t:=\exp(-\rmi\mf{H}_{\text{SB}}t)$ is the time evolution operator (Stone theorem~\cite{simon}). Needless to say, such a form
of the solution is useless for practical purposes.

There is at least one important reason for which a manageable form of the time evolution operator $\mf{U}_t$ is worth seeking for.
Namely, it allows to construct the exact reduced time evolution of the spin immersed within the bosonic bath, the so-called reduced
dynamics~\cite{Alicki}:

\begin{equation}
  \label{reduced}
 \rho_t = \text{Tr}_{\text{B}}(\mf{U}_t\rho_0\otimes\omega_{\txt{B}}\mf{U}_t^{\dagger}).
\end{equation}
Above, the state $\omega_{\txt{B}}$ is an initial state of the bosonic bath. $\txt{Tr}_{\txt{B}}$ denotes the partial trace, \ie, 
$\txt{Tr}_{\txt{B}}(\txt{M}\otimes X)=\txt{M}\txt{Tr}X$, where $\mbox{Tr}$ refers to the usual trace on $\mc{F}_{B}$. For the
sake of simplicity we have assumed that the initial state of the composite system $\rho_{\txt{int}}$ is the tensor product of
the states $\rho_0$ and $\omega_{\txt{B}}$. In other words, no initial correlations between the systems are 
present~\cite{korelacje,erratum,KrausRep,dajka3}.

In general, the formula~(\ref{reduced}) is far less useful, than its theoretical simplicity might indicate. Indeed, to trace out
the state $\mf{U}_t\rho_{\txt{int}}\mf{U}_t^{\dagger}$ over the bosonic degrees of freedom, one needs to i) calculate $\mf{U}_t$
and ii) apply the result to the initial state $\rho_{\text{int}}$. Herein we will cover the first step and we will investigate
the ability to accomplish the second one.

In order to write the time evolution operator $\mf{U}_t$ in a computationally accessible form, the diagonalization of its generator
$\mf{H}_{\text{SB}}$ or an appropriate factorization~\cite{faktoryzacja} is required. It can be found (see \eg,~\cite{mgr,RiccEq,gardas2})
that the problem of diagonalization on the Hilbert space $\mb{C}^{2}\ot\mc{F}_{\text{B}}$ can be mapped to the problem of resolving the
Riccati equation~\cite{Vadim}. This new approach was recently successfully applied to the time-dependent spin-spins model~\cite{gardas3}.
As a result, the exact reduced dynamics of the qubit in contact with a spin environment and in the presence of a precessing magnetic
field has been obtained. It is interesting, therefore, to apply this approach to the SB model as well. This paper has been devoted to
accomplish this purpose. Although, an explicit form of the Riccati equation has already been derived~\cite{gardas}, the solution has not
been provided yet. In this paper we derive an exact solution of this equation assuming $\beta=0$.

\section{The block operator matrix representation and the Riccati equation}

We begin by reviewing some basic facts concerning a connection between the theory of block operator matrices~\cite{spectral} and the
SB model. First, the Hamiltonian~(\ref{eq:SB}) admits the block operator matrix representation~\cite{gardas,bom}:

\begin{equation}
\label{eq:bomSB}
\mf{H}_{\text{SB}} =
   \begin{bmatrix}
    \txt{H}_{\text{B}}+\txt{V}+\beta & \alpha \\
    \alpha & \txt{H}_{\text{B}}-\txt{V}-\beta
    \end{bmatrix}
    \equiv
    \begin{bmatrix}
    \txt{H}_{\p} & \alpha \\
    \alpha & \txt{H}_{\m}
     \end{bmatrix},
    %\quad\text{where}\text\quad
    %\txt{H}_{\spm} = \txt{H}_{\text{B}}\pm \txt{V}\pm\beta,
\end{equation}
with respect to the direct sum decomposition $\mc{H}_{\text{tot}}=\mc{F}_{\text{B}}\oplus\mc{F}_{\text{B}}$ of $\mc{H}_{\text{tot}}$.
The entries $\alpha$ and $\beta$ of the operator matrix~(\ref{eq:bomSB}) are understood as $\alpha\Id{B}$ and $\beta\Id{B}$, respectively.
Henceforward, we use the same abbreviation for any complex number.

The Riccati operator equation associated with the matrix~\eref{eq:bomSB} reads~\cite{gardas}

\begin{equation}
\label{eq:riccatiSB}
 \alpha X^{2} + X\txt{H}_{\p}-\txt{H}_{\m}X-\alpha = 0,
\end{equation}
where $X$ is an unknown operator, acting on $\mc{F}_{\txt{B}}$, which needs to be determined. The solution of this equation, if it
exists, can be used to diagonalize the Hamiltonian~\eref{eq:bomSB}. To be more specific, if $X$ solves~(\ref{eq:riccatiSB}) the 
following equality holds true

\begin{equation}
\label{eq:diag}
\mf{S}^{-1}\mf{H}_{\text{SB}}\mf{S} =
       \begin{bmatrix}
    \txt{H}_{\p}+\alpha X & 0 \\
    0 & \txt{H}_{\m}-\alpha X^{\dg}
    \end{bmatrix},
    \quad\text{where}\quad
     \mf{S} =
    \begin{bmatrix}
    1 & -X^{\dg} \\
    X & 1
    \end{bmatrix}.
\end{equation}
By means of this decomposition we can write $\mf{U}_t$ in an explicit matrix form: 

\begin{equation}
\label{eq:evolve}
\mf{U}_t = \mf{S}\mbox{diag}[e^{-\rmi\left(\txt{H}_{\p}+\alpha X\right)t}, e^{-\rmi\left(\txt{H}_{\m}-\alpha 
X^{\dagger}\right)t}]\mf{S}^{-1}.
\end{equation}
Note, the last formula reduces the problem of finding the solution of the Schr\"{o}dinger equation~\eref{eq:Sch} to the problem of resolving
the Riccati equation~\eref{eq:riccatiSB}. It is well-established that the reduced dynamics~\eref{reduced} can easily be obtained when
$\alpha=0$~\cite{SB_Alicki}. In this case no additional assumptions on $\beta$ are needed, which should not be surprised since the
matrix~\eref{eq:bomSB} is already in a diagonal form ($X=0$). Moreover, if $\alpha=0$ the qubit does not exchange the energy with the bosonic
field because $[\txt{H}_{\text{S}}\ot\Id{B}, \mf{H}_{\text{SB}}]=0$. Therefore, the only exactly solvable case, which is known at the present time,
represents rather extreme physical situation.

In the next section we will derive an exact solution of the RE~\eref{eq:riccatiSB} assuming $\beta=0$; nevertheless, we not impose any
restrictions on $\alpha$. This is exactly the opposite situation to the one we have discussed above. At this point, the natural question
can be addressed: what about the case, when both $\alpha$ and $\beta$ are not equal to zero? Unfortunately, the answer is still to be 
found. In fact, usually the SB model is defined only for $\beta=0$. At first, it might seem that the complexity of the problem is the 
same both for $\beta=0$ and $\beta\not=0$. Although, this is indeed true when $\alpha= 0$, no argument proving this conjecture for 
$\alpha\not=0$ has been given so far. We will return to this matter at the end of the next section. 

\section{Solution of the Riccati equation}
  \subsection{Single boson case}
To understand the idea of our approach better let us first consider the case where there is only one boson in the bath~\cite{CJC,JC}.
Then, the Hamiltonian of the SB model can be written by using the block operator matrix nomenclature as ($\beta=0$)

\begin{equation}
 \label{eq:singleBOM}
\mf{H}_{\txt{SB}} =
  \begin{bmatrix}
    \txt{H}_{\m} & \alpha \\
   \alpha & \txt{H}_{\p}
   \end{bmatrix}
   \quad\text{with}\quad
   \txt{H}_{\spm} = \omega a^{\dagger}a\pm(g^{\ast}a+ga^{\dagger}).
\end{equation}
The operators $\txt{H}_{\spm}$ can be expressed in a more compact form, that is

\begin{equation}
  \label{eq:rules}
\txt{H}_{\m} = \omega\txt{D}_fa^{\dagger}a\txt{D}_{\m f}- E
\quad\text{and}\quad
\txt{H}_{\p} = \omega\txt{D}_{\m f}a^{\dagger}a\txt{D}_f- E,
\end{equation}
where $f=g/\omega$ and $E=|g|^2/\omega$. The displacement operator $\txt{D}_f := \exp(f^*a-fa^{\dg})$
has the following, easy to prove, properties

\begin{equation}
\label{eq: weyl}
\txt{i)}\quad\txt{D}_{-f}=\txt{D}_f^{\dg},\quad\text{ii)}\quad \txt{D}_f\txt{D}_{\m f}=\Id{B}
\quad\txt{and}\quad\text{iii)}\quad
\txt{D}_{f}\txt{D}_{g} = e^{\rmi\Im (fg^*)}\txt{D}_{f+g}.
\end{equation}
$\Im$ stands for the imaginary part of the complex number $fg^*$. The relations~\eref{eq:rules} can be proven by using equality
$\txt{D}_fa\txt{D}_{\m f} = a-f$, which follows from the Baker-Campbell-Hausdorff formula~\cite{galindo1,BCH2}. For the sake of
simplicity and without essential loss of generality we rescale the Hamiltonian~\eref{eq:singleBOM} so that 
$\mf{H}_{\text{SB}}\to\mf{H}_{\text{SB}}+E$. This is nothing but a rescaling of the reference point of the total. 

After this procedure the Hamiltonian~\eref{eq:singleBOM} takes the form

\begin{equation}
\label{eq:rescal}
\mf{H}_{\txt{SB}} =
  \begin{bmatrix}
   \omega\txt{D}_fa^{\dagger}a\txt{D}_{\m f}  & \alpha \\
   \alpha & \omega\txt{D}_{\m f}a^{\dagger}a\txt{D}_f
   \end{bmatrix},
\end{equation}
while the corresponding Riccati equation reads

\begin{equation}
\label{eq:riccatiSB2}
 \alpha X^{2} + X\left(\omega\txt{D}_fa^{\dagger}a\txt{D}_{\m f}\right)-
               \left(\omega\txt{D}_{\m f}a^{\dagger}a\txt{D}_f\right)X-\alpha = 0.
\end{equation}
To solve this equation, let us first define an operator $\txt{P}_{\varphi}$ in a way that

\begin{equation}
\label{eq:Pdef}
\txt{P}_{\varphi} := \exp (\rmi\varphi a^{\dg}a), \quad \varphi\in[0,2\pi).
\end{equation}
%
%From this definition follows that
It is not difficult to see that

\begin{equation}
 \label{eq:prop}
 \txt{i)}\quad\txt{P}_{\m\varphi}=\txt{P}_{\varphi}^{\dg},
 \quad\txt{ii)}\quad
 \txt{P}_{\varphi}\txt{P}_{\m\varphi}=\Id{B}
 \quad\text{and}\quad\txt{iii)}\quad
  \txt{P}_{\varphi}\txt{P}_{\psi} = \txt{P}_{\varphi\p\psi}.
\end{equation}
%
%Since $\txt{P}_{\varphi}$ is a function of $a^{\dagger}a$, we instantly obtain $[\txt{P}_{\varphi},a^{\dagger}a]=0$.
Moreover, from the Baker-Campbell-Hausdorff formula we also have $\txt{P}_{\varphi}a\txt{P}_{\m\varphi} = e^{-\rmi\varphi}a$,
which ultimately leads to

\begin{equation}
\label{eq:trans}
 \txt{P}_{\varphi}\txt{D}_f\txt{P}_{\m\varphi} = \txt{D}_{e^{\rmi\varphi}f}.
\end{equation}

In what follows, we will prove that $\txt{P}_{\pi}$ solves the Riccati equation~(\ref{eq:riccatiSB2}). First, let us note that 
$\txt{P}_{\pi}$ is a function of the number operator $a^{\dagger}a$, thus $[\txt{P}_{\pi},a^{\dagger}a]=0$. In view of~\eref{eq:trans}
we obtain $\txt{P}_{\pi}\txt{D}_f\txt{P}_{\m\pi} = \txt{D}_{\m f}$, hence

\begin{equation}
\label{eq:unit}
\txt{P}_{\pi}\left(\txt{D}_fa^{\dg}a\txt{D}_{\m f}\right) =
\left(\txt{D}_{\m f}a^{\dg}a\txt{D}_f\right)\txt{P}_{\pi}.
\end{equation}
By writing $\txt{P}_{\pi}$ in terms of the eigenstates $\ket{n}$ of $a^{\dg}a$ we obtain
%Second, let us rewrite $\txt{P}_{\pi}$ in terms of the eigenstates $\ket{n}$ of $a^{\dg}a$. By doing so we obtain

\begin{equation}
\label{eq:parity}
 \txt{P}_{\pi} = \sum_{n\in\mb{N}}e^{i\pi n}\ket{n}\bra{n}
               = \sum_{n\in\mb{N}}(-1)^n\ket{n}\bra{n},
\end{equation}
where we used the well-known mathematical fact that $a^{\dg}a\ket{n}=n\ket{n}$, for $n\in\mb{N}$. Finally, from~(\ref{eq:parity}) 
we conclude that $\txt{P}_{\pi}$ is an involution, \ie, $\txt{P}_{\pi}^{2}=\Id{B}$, which together with~\eref{eq:unit} leads to

\begin{equation}
\label{eq:sol}
  \alpha\txt{P}_{\pi}^{2}+  \txt{P}_{\pi}\left(\omega\txt{D}_fa^{\dg}a\txt{D}_{\m f}\right)-
                            \left(\omega\txt{D}_{\m f}a^{\dg}a\txt{D}_f\right)\txt{P}_{\pi}-\alpha = 0.
\end{equation}
Note, $\txt{P}_{\pi}$ transforms the creation $a^{\dg}$ and annihilation $a$ operators into $-a^{\dg}$ and $-a$, respectively.
In other words, $\txt{P}_{\pi}$ can be interpreted as the bosonic parity operator~\cite{Bender}. Moreover, $\txt{P}_{\pi}$
does not depend on the parameter $\alpha$; in particular, $\txt{P}_{\pi}$ remains a nontrivial ($X\not=0$) solution of the 
Riccati equation~\eref{eq:riccatiSB2} even when $\alpha=0$ (Sylvester equation).

Now, by means of the parity operator $\text{P}_{\pi}$, we can derive an accessible form of the time evolution operator $\mf{U}_t$.
According to~\eref{eq:diag} and~\eref{eq:evolve} we have

\begin{equation}
\label{eq:esolve}
\mf{U}_{t} =\frac{1}{2}
    \begin{bmatrix}
    \text{U}_{\p}(t) & \txt{V}_{\p}(t)\txt{P}_{\pi} \\
    \txt{V}_{\m}(t)\txt{P}_{\pi} & \txt{U}_{\m}(t)
   \end{bmatrix},
\end{equation}
where the quantities $\txt{U}_{\spm}(t)$ and $\txt{V}_{\spm}(t)$ read as follows

\begin{equation}
  \txt{U}_{\spm}(t) = e^{-\rmi(\txt{H}_{\spm}+\alpha\parity)t} + e^{-\rmi(\txt{H}_{\spm}-\alpha\parity)t},
  \quad
  \txt{V}_{\spm}(t) = e^{-\rmi(\txt{H}_{\spm}+\alpha\parity)t} - e^{-\rmi(\txt{H}_{\spm}-\alpha\parity)t}.
\end{equation}
For $\alpha=0$ the formula~\eref{eq:esolve} simplifies to the well-known result~\cite{SB_Alicki}, which can be obtained
independently, without solving the Riccati equation.

It is instructive to see how the bosonic parity operator $\txt{P}_{\pi}$ can also be used to construct the constant of motion for
the SB model. For this purpose let us take $\mf{J}_{\pi}:=\sigma_x\otimes\txt{P}_{\pi}$; then $[\mf{J}_{\pi},\mf{H}_{\txt{SB}}]=0$,
thus from the Heisenberg equations of motion follows $\dot{\mf{J}}_{\pi}=0$, which means that $\mf{J}_{\pi}$ does not vary with 
time. Since $\txt{P}_{\pi}$ is an involution, \ie, $\txt{P}_{\pi}^2=\Id{B}$ thus $\mf{J}_{\pi}$ is an involution as well. Therefore,
$\mf{J}_{\pi}$ can be seen as the parity operator of the total system. In conclusion, the total parity is conserved when $\beta=0$.

For $\beta\not=0$ the parity symmetry of the total system is broken and the Riccati equation~(\ref{eq:riccatiSB2}) cannot be solved
by applying a similar method to the one we have used above in the case of $\beta=0$. From mathematical point of view, the problem 
arises because the diagonal entries $\txt{H}_{\txt{B}}\pm V\pm\beta$ are no longer related by an unitary transformation. Indeed, if
the converse would be true, there would then exist an unitary operator $\txt{W}$ such that $\txt{W}^{\dg}\left(\txt{H}_{\txt{B}}+ 
V+\beta\right)\txt{W} =\txt{H}_{\txt{B}}-V-\beta$. Thereby the spectra $\sigma\left(\txt{H}_{\txt{B}}\pm V\pm\beta\right)=
\sigma(\txt{H}_{\txt{B}}\pm V)\cup\{\pm\beta\}$ would be the same, which clearly is impossible unless $\beta=0$. As a result, for
$\alpha\not=0$ one can expect that the mathematical complexity of the SB model is different within the regimes $\beta=0$ and $\beta\not=0$.

\subsection{Generalization}

The results of the preceding subsection can be generalized to the case where there is more that one boson in the bath. In order to 
achieve this objective one needs to redefine the displacement operator $\txt{D}_f$ in the following way

\begin{equation}
\label{eq:Rweyl}
\txt{D}_f\rightarrow\exp\left(A-A^{\dg}\right),
\quad\text{where}\quad
A = \sum_{k} \frac{g_k^*}{\omega_k}a_k.
\end{equation}
Then, the solution of the Riccati equation reads

\begin{equation}
\label{eq:Rparity}
 \txt{P} = \exp(\rmi\pi\sum_ka_k^{\dg}a_k)
 =\bigotimes_{k}\txt{P}_{\pi,k},
 \quad\text{where}\quad \txt{P}_{\pi,k} = \exp(i\pi a_k^{\dagger}a_k).
\end{equation}
%where $\txt{P}_{\pi,k} = \exp(i\pi a_k^{\dagger}a_k)$. %is the parity operator for

\section{Remarks and Summary}

In this article, we have solved the Riccati operator equation associated with the Hamiltonian
of the paradigmatic spin-boson model. Next, in terms of the solution we have derived an explicit matrix
form of the time evolution operator of the total system. This, in particular, allows us to solve the 
Schr\"{o}dinger equation~(\ref{eq:Sch}). We wish to emphasize that in order to obtain the reduced 
dynamics~(\ref{reduced}) one more step is required. Namely, the terms like

\begin{equation}
 \label{trace}
\mbox{Tr}(e^{-i[\txt{H}_{\spm}\pm\alpha\txt{P}_{\pi}]t}\omega_{\txt{B}}e^{i[\txt{H}_{\spm}\pm\alpha\txt{P}_{\pi}]t})
\end{equation}
need to be determined.
Of course, one can always evaluate the quantities given above by using \eg, perturbation
theory. However, the true challenge is to establish this goal without approximations.
It seems that the simplest way to do so is to solve the eigenvalue problem 
$(\txt{H}_{\spm}\pm\alpha\txt{P}_{\pi})\ket{\psi}=\lambda\ket{\psi}$. The ability to solve
this eigenproblem separates us from deriving the exact reduced dynamics of the qubit immersed
within the bosonic bath. We stress that for $\alpha\not=0$ the problem is nontrivial since the
qubit exchange the energy with its environment. Moreover, an impact on the mathematical complexity
of the model, has not only a transfer of the energy between the systems but also the energy
split between the states $\ket{0}$, $\ket{1}$.

Interestingly, the Riccati equation is a second order operator equation, thus one can expect
that its solution involves a square root. In particular, nothing indicates that the solution should be
linear as it is in our case. Therefore, we not only solved the Riccati equation~(\ref{eq:riccatiSB})
but also linearized the solution. At this point, a worthwhile question can be posed: is it a
coincidence that the linear operator happens to solve a nonlinear equation? Perhaps, it is a
manifestation of some additional structure in the model. Historically, a similar situation took place
when Dirac solved the problem with a negative probability by introducing His famous equation~\cite{dirac}.
By linearizing the Hamiltonian of the relativistic electron, Dirac not only predicted an existence of
the antiparticles but also explained the origin of the additional degree of freedom of the electron.
\ack
%\begin{acknowledgments}
The author would like to thank Jerzy Dajka for helpful comments and suggestions.
%discussions.
%\end{acknowledgments}

\section*{References}

 %\bibliographystyle{jphysicsB}
%\bibliographystyle{unsrt}
%\bibliography{spin_boson}
%
            %spin_boson.bbl%
%
 
\end{document}